\documentstyle[preprint,pre,aps]{revtex}
\begin{document}
\draft
\title{Reentrant phase diagram of branching annihilating
      random walks with one and two offsprings}
\author{Sungchul Kwon and Hyunggyu Park}
\address{Department of Physics, Inha University,
             Inchon 402-751, Korea}
\date{\today}
\maketitle
\begin{abstract}
We investigate the phase diagram of branching annihilating random
walks with one and two offsprings in one dimension. A walker
can hop to a nearest neighbor site or branch with one or two
offsprings with relative ratio. Two walkers annihilate immediately
when they meet. In general, this model exhibits a continuous phase
transition from an active state into the absorbing state (vacuum) at a
finite hopping probability. We map out the phase diagram by Monte
Carlo simulations which shows a reentrant phase transition from
vacuum to an active state and finally into vacuum again as the
relative rate of the two-offspring branching process increases. This
reentrant property apparently contradicts the conventional wisdom
that increasing the number of offsprings will tend to make the system
more active. We show that the reentrant property is due to the
static reflection symmetry of two-offspring branching processes and the
conventional wisdom is recovered when the dynamic reflection symmetry
is introduced instead of the static one.
\end{abstract}

\pacs{PACS numbers: 64.60.-i, 02.50.-r, 05.70.Ln, 82.65.Jv}

\section{Introduction}

In recent years, various kinds of nonequilibrium lattice models
exhibiting a continuous phase transition from a reactive phase
into an absorbing (inactive) phase have been studied extensively
\cite{Liggett,Dickb}.
Most of models investigated are found to belong to the directed
percolation (DP) universality class \cite{Jan,Grass82,Cardy,Grin}.
A common feature of these
models is that the absorbing phase consists of a {\em single}
absorbing state. Even some models with infinitely many absorbing
states also exhibit critical behavior in the DP universality class
\cite{Jen932,Jen933}.

Only a few models have been studied which are not in the DP
universality class. Those are the model
A and B of probabilistic cellular automata \cite{Grass84,Grass89},
nonequilibrium kinetic Ising models with two different dynamics
\cite{Meny94,Meny95}, and interacting monomer-dimer models
\cite{Park94,Park951,Park952}. Numerical
investigations show that they belong to the same but non-DP
universality class.
These models share a common property that the absorbing phase
consists of two {\em equivalent} absorbing states.

Recently, the branching annihilating random walks (BAW)
have been studied intensively \cite{Bram,Sud,Taka,Jen934,Jen931,Jen941}.
The BAW model is a lattice model where
a walker can hop to a nearest neighbor site with probability $p$
and branch with $n$ offsprings in nearest neighborhood
with probability $1-p$. Two walkers
annihilate immediately when they meet. In general, this model
exhibits a continuous phase transition from an active state
into the absorbing state (vacuum) at finite hopping probability
$p_c$.
Even though the BAW model has a single
absorbing state, its critical property depends on the parity of the
number of offsprings created in a branching process.
Dynamics of the BAW models with even $n$ conserve the number of walkers
modulo 2, while the BAW models with odd $n$ evolve without any conservation.
The BAW models with odd $n$ exhibit the DP critical behavior
\cite{Taka,Jen934}, while
the BAW models with even $n$ exhibit the same non-DP behavior
as in the models with two equivalent absorbing states
\cite{Grass89,Park94,Park951,Park952,Jen941}.
One can find that the total number of kinks are conserved modulo 2
in models with two equivalent absorbing states. Therefore this conservation
law may be responsible for the non-DP critical behavior.

The critical exponents characterizing the non-DP behavior
are measured accurately by extensive Monte Carlo simulations for the BAW
model with $n=4$ in one dimension \cite{Jen941}.
Unfortunately, the BAW model with $n=2$ which is simpler
does not have an active phase ($p_c=0$) \cite{Sud,Taka2}.
Recently, ben-Avraham {\em et al} \cite{Redner}
introduce another parameter which controls the two-walker
annihilation process and show that the BAW model with $n=2$
exhibits a continuous phase transition at a finite value of
$p_c$ except for the case of infinite annihilation rate (the ordinary BAW
model). But the values of the critical exponents are not reported there.
As the annihilation rate becomes smaller, it is clear that
the system gets more active (more walkers survive) so the critical
probability $p_c$ goes higher.

In the ordinary BAW models, there are two competing elementary processes,
i.e.~hopping and branching. The hopping process does not increase
the number of walkers but can decrease it by two-walker annihilations.
As the hopping probability becomes larger, the system tends to
trap into the absorbing state (vacuum). The branching process may
increase the number of walkers so it can make the system more active.
That is why the BAW model with one offspring exhibits a phase transition
from an active state into vacuum as $p$ becomes larger.
Therefore one may argue that increasing the number of offsprings in the
branching process will make the system more active so the
critical probability $p_c$ will go higher as $n$ becomes bigger.
However, numerical and analytical study of the BAW models with $n$ offsprings
show that $p_c$ is not a monotonically increasing function of $n$.
In fact, the values of $p_c$ are 0.1070(5), 0, 0.459(1), 0.7215(5),
and 0.718(1) for $n=1-5$ \cite{Taka,Jen934,Jen941}.
These results contradict the conventional
wisdom mentioned above and there is no consistent manner in making
the system more active by changing the number of offsprings.

In order to understand this rather surprising result, we introduce
a model which interpolates the BAW models with one offspring
and with two offsprings.
In this model, the branching process creates one offspring or
two offsprings with relative ratio. We map out the phase diagram
by dynamic Monte Carlo simulations which shows a reentrant phase
transition from vacuum to an active state and finally into
vacuum again as the relative rate of the two-offspring process increases
at fixed hopping probability. The second phase transition occurs at
quite high rates of the two-offspring process (more than $80\%$).
We argue that the second transition is due to the static reflection
symmetry of the two-offspring branching process (one offspring
to the left and the other to the right of the branching walker:
{\em static branching}).
This reentrant second transition disappears and
our conventional wisdom is recovered when the dynamic reflection
symmetry is introduced (both offsprings to the left or to the right
of the branching walker with equal probability: {\em dynamic branching}).

In the next section, we describe the BAW model with one and two
offsprings with relative ratio. Dynamic Monte Carlo results
are discussed which show the reentrant phase diagram.
In Sec.~III, the BAW model with dynamic branching
is introduced. Our numerical results for $n=2$ find the existence
of the non-DP critical behavior at finite hopping probability.
In Sec.~IV, we study the BAW model with one and two offsprings
created by dynamic branching. The reentrant
behavior disappears entirely as expected. We conclude in Sec.~V
with brief summary.

\section{The BAW model with one and two offsprings: Static branching}

We consider the BAW model with one and two offsprings with relative
ratio in one dimension. The evolution rules of this model are given
as follows. First, choose a walker at random. It may hop to
a randomly chosen nearest neighbor site with probability $p$.
If this site is already occupied by another walker, both walkers
annihilate immediately. With probability $1-p$, the randomly chosen
walker creates two offsprings symmetrically at nearest neighbor sites
with relative probability $r$ ($0\le r \le 1$) or
creates one offspring at a randomly chosen nearest neighbor
site with relative probability $1-r$.
When an offspring is created on a site already occupied, both
walkers annihilate immediately.
The case $r=0$ corresponds
to the BAW model with one offspring which exhibits
a continuous phase transition from an active phase into vacuum
at $p\simeq 0.1070$ \cite{Jen934}.
The other limiting case $r=1$ corresponds to
the BAW model with two offsprings which does not have an active
state at finite values of $p$ \cite{Sud,Taka2}.

We perform dynamic Monte Carlo simulations for this model
with various values of $r=0$, 0.25, 0.50, 0.75, 0.8, 0.85, 0.9,
0.95, and 1.
We start with two nearest neighbor
walkers at the central sites of a lattice. Then the system evolves
along the dynamical rules of the model.
After one attempt of hopping or branching on the average per lattice
site (one Monte Carlo step), the time is incremented by one unit.
A numer of independent runs, typically $10^5$, are made up to
2000 time steps for various values of $p$ near the critical
probability $p_c$. Most runs, however, stop earlier because the
system gets into the absorbing state.
We measure the survival probability $P(t)$ (the
probability that the system is still active at time $t$),
the number of walkers $N(t)$ averaged over all
runs, and the mean-square distance of spreading $R^2 (t)$ averaged
over the surviving runs.
At criticality, the values of these quantities scale
algebraically in the long time limit \cite{Grass79}
\begin{eqnarray}
P(t) &\sim& t^{-\delta},\nonumber \\
N(t) &\sim& t^{\eta},\\
R^2(t) &\sim& t^{z},\nonumber
\end{eqnarray}
and double-logarithmic plots of these values against time
show straight lines. Off criticality, these plots show
some curvatures.
More precise estimates for the scaling exponents can be obtained
by examining the local slopes of the curves.
The effective
exponent $\delta(t)$ is defined as
\begin{equation}
-\delta(t) = \frac{\log \left[ P(t) / P(t/b) \right]}{\log ~b}
\end{equation}
and similarly for $\eta (t)$ and $z(t)$. In Fig.~1, we plot
the effective exponents against $1/t$ with $b =5$ for $r=0.25$.
Off criticality these plots
show upward or downward curvatures. From Fig.~1, we estimate
$p_c = 0.1265(5)$ and dynamic exponents
$ \delta = 0.158(2)$, $\eta = 0.310(3)$, $z = 1.27(1)$.
These values are in an excellent accord with those of the
DP universality class;  $\delta=0.1596(4),
\eta=0.3137(10)$, and $z=1.2660(14)$ (see reference \cite{Jen941}).
This result supports the conjecture that models with a single
absorbing state and no conservation laws should belong to the DP
universality class.

Similarly we determine the values of the critical probability $p_c$
and the dynamic exponents for various values of $r$. As expected
from the above conjecture, the values
of the dynamic exponents stay almost unchanged except for $r=1$.
Estimates of $p_c$ for various values of $r$ are listed in Table I.
The value of $p_c$ slowly increases as $r$ varies upto $\sim 0.75$ and
drastically decreases to zero as $r$ approaches the value of unity.
The $r-p$ phase diagram is drawn in Fig.~2 where a reentrant
phase transition is explicitly shown. At fixed values of
$p=0.11\sim 0.14$, the system undergoes phase transitions from
vacuum to an active state and finally into vacuum again as $r$
becomes larger (more offsprings are created).
The second transition occurs at quite high rates of
the two-offspring branching process.
This result implies that our conventional wisdom does not
apply near $r=1$.

The BAW model with two-offsprings can be solved exactly
at $p=0$, where pairs of nearest neighbor walkers diffuse
like the ordinary random walks, i.e.~$\cdots 001100\cdots \rightarrow
\cdots 000110\cdots$ where `$1$' represents an occupied site and `$0$'
a vacant site.
When two pairs collide each other, they just bounce back.
So the number of walkers is bounded in this model, in contrast
to the BAW model with one offspring where the number of walkers
is not bounded. So our conventional wisdom that the BAW model
with more offsprings may be more active does not work.
For $p>0$, these pairs annihilate by hopping processes so the system
becomes inactive in the long time limit.

We notice that picture of diffusing pairs
is mainly due to the static reflection symmetry of the branching
process. If the dynamic reflection symmetry is adopted instead of
the static one,
this diffusing-pair picture is no longer valid. In this case,
a walker branches two offsprings both to the left or to the
right of itself with equal probability. The reflection
symmetry is not broken on average, but this branching
process allows a formation of long chains of walkers,
i.e.~$\cdots 00011000 \cdots \rightarrow
\cdots 01111000\cdots$. So
the number of walkers is not bounded from above and
an active phase may appear at finite hopping probability.

\section{The BAW model with two offsprings: Dynamic branching}

We study the BAW model with two offsprings created by
dynamic branching. The hopping process is the same as
in the ordinary BAW models, but in the branching process
a randomly chosen walker creates two
offsprings both on the sites to the left or to the right of the walker
with equal probability.

We perform dynamic Monte Carlo simulations, starting
from a pair of walkers at the central sites of a lattice.
$10^6$ independent runs are made during 2000 time steps
and we measure $P(t)$, $N(t)$, and $R^2 (t)$.
In Fig.~3, we plot the effective exponents $\delta(t)$,
$\eta(t)$, and $z(t)$ against $1/t$ with $b=5$.
These plots clearly show the existence of an active phase.
We estimate $p_c=0.5105(7)$ and the dynamic exponents
$\delta=0.287(1)$, $\eta=0.000(3)$, and $z=1.155(5)$.
These values are in an excellent accord with those of
the non-DP universality class; $\delta=0.285(2)$, $\eta=0.000(1)$,
and $z=1.141(2)$ for the BAW model with four offsprings.
This result supports the conjecture that models with
particle number conservation of modulo 2 should belong
to the same but non-DP universality class.

We also perform dynamic simulations with a different
initial configuration. We start with a single walker at
the center of a lattice. Conservation of the number of
walkers of modulo 2 prevents the system
from entering the absorbing state (vacuum). So the survival
probability exponent $\delta$ must be zero. Our numerical results
conclude that the dynamic exponents $\eta=0.283(5)$ and
$z=1.15(1)$. These values also agree very well with those of
the BAW model with four offsprings ($\eta=0.282(4)$, $z=1.140(5)$).

Compared with the ordinary BAW model with two offsprings, it is clear
that the static reflection symmetry is responsible for the
nonexistence of an active phase and the dynamic reflection symmetry
makes the system more active. To support this idea, we employ the
mean field theory on spreading of the active region.
Difference between these two models lies on the branching process.
So we only consider the effect of different branching mechanism
on the boundary of the active region. The boundary can move by
branching of walkers at the boundary or nearby. For example,
a branching process of a walker at the boundary increases the
active region by one unit in the BAW model with the static symmetry.
For the BAW model with the dynamic symmetry, the same process
increases the active region by the same amount (one unit on average).
However, a branching process of a walker near the boundary inside
of the active region decreases the active region
differently for these two models.
Considering all possible cases that the boundary can move and
using the mean field theory to assign a proper probability to
each case, we find the outward velocity of the boundary as
\begin{eqnarray}
v_s &=& (1-p)\rho (1-\rho), \nonumber\\
v_d &=& (1-p)\rho (1-\frac{\rho^2}{2}),
\end{eqnarray}
where $v_s$ ($v_d$) is the outward velocity of the boundary for the
model with the static (dynamic) symmetry and $\rho$ is the
walker density. The first term comes from the branching
process of a walker at the boundary and the second term near the
boundary. The effect of hopping on the boundary velocity is omitted.
We find that $v_s$ is always smaller than $v_d$.
This result implies that the active region grows faster in the
model with the dynamic symmetry, so this model should be more
active than the model with the static symmetry. This mean field
argument can be easily generalized to the models with general $n$
offsprings.

As the critical probability depends enormously on the symmetry
of branching processes (especially $n=2$), it is meaningless
to ask whether $p_c$ monotonically increases with $n$ for
the ordinary BAW models. These models have not
been classified by the symmetry of branching processes.
The branching process of the
ordinary BAW model with one offspring basically belongs to
the process with the dynamic symmetry. If we compare
the critical probabilities of the BAW models with dynamic
branching only, we expect that $p_c$ will monotonically increase with
$n$, i.e.~the dynamic branching process always makes the system
more active. In the next section, we study the BAW model
with one and two offsprings created by dynamic branching
and examine whether the reentrant behavior seen in the case
of static branching (Sec.~II) disappears.

\section{The BAW model with one and two offsprings: Dynamic branching}

We consider the BAW model with one and two offsprings branched
dynamically with relative ratio. The evolution rules of this model
are exactly the same as in Sec.~2 except that static branching
is replaced with dynamic branching for the two-offspring branching
process.

We perform dynamic Monte Carlo simulations for $r=0.25$, 0.50, 0.75.
Estimated values of $p_c$ are listed in Table II. As expected,
$p_c$ monotonically increases as $r$ becomes larger (more offsprings
are created). The reentrant phase transition disappears entirely
in this model with dynamic branching (see Fig.~4).
Of course, the critical behavior at the absorbing
transitions belongs to the DP universality class for $0\le r <1$.

\section{Summary}

We study the BAW models with static (ordinary)
branching and dynamic branching.
With the static branching, the BAW model with
one and two offsprings shows a reentrant phase transition from
vacuum to an active state and finally into vacuum again as the
relative rate of the two offspring process increases. We argue that
this reentrant property originates from the static reflection
symmetry of the two-offspring branching process.

The ordinary BAW model with two offsprings does not have an active
phase at finite values of hopping
probability. We introduce the BAW model with two offsprings created
by dynamic branching and show that this model
exhibits a continuous phase transition from an active phase into vacuum
at finite hopping probability. Its critical behavior belongs to
the same non-DP universality class as in the ordinary BAW model with four
offsprings. We also study the BAW model with one and two offsprings
created by dynamic branching. The reentrant behavior disappears
and our conventional wisdom is recovered as
expected. Our results shed light on how the system can be active
by different branching processes and the BAW model with dynamic
branching may serve as another simple model exhibiting the non-DP
critical behavior.

\section*{acknowledgements}

We wish to thank Heungwon Park for interesting
discussions. This work is supported in part by the BSRI, Ministry of
Education (Grant No.~95-2409) and by an Inha University research
grant (1995).

\newpage
\begin{center}
{\Large {\bf Tables}}
\vspace{5mm}
\end{center}
\begin{description}
\item[{\bf Table I :}]
Critical hopping probability $p_c$ for the BAW model
with one and two offsprings created by the static branching.
$r$ is the relative probability of the two-offspring process.
Numbers in parentheses represent the errors in the last digits.
\end{description}
\begin{center}
\begin{tabular}{cccccccccc}
\hline\hline
$r$ \ & 0 & 0.25 & 0.50 & 0.75 & 0.80 & 0.85 & 0.90 & 0.95 & 1 \\ \hline
$p_c$\   & 0.1070(5)\  & 0.1265(5)\   & 0.1415(5)\   & 0.1480(5)\
& 0.1470(5)\   & 0.1430(5)\  & 0.1380(5)\  & 0.1250(5)\  & 0
\\ \hline\hline
\end{tabular}
\end{center}
\vspace{5mm}
\begin{description}
\item[{\bf Table II :}]
Critical hopping probability $p_c$ for the BAW model
with one and two offsprings created by the dynamic branching.
$r$ is the relative probability of the two-offspring process.
Numbers in parentheses represent the errors in the last digits.
\end{description}
\begin{center}
\begin{tabular}{cccccc}
\hline\hline
$r$\ & 0 & 0.25 & 0.50 & 0.75 & 1 \\ \hline
$p_c$\ & 0.1070(5)\ & 0.18(1)\ & 0.25(1)\ & 0.315(5)\
& 0.5105(7)\ \\ \hline\hline
\end{tabular}
\end{center}

\newpage
{\center\Large Figure Captions}
\begin{description}
   \item[Fig.\  1  :]  Plots of the effective exponents against
    $1/t$. Three curves from top to bottom
    in each panel correspond to $p=0.1260$, 0.1265, 0.1270.
    Thick lines are critical lines ($p=0.1265$).

   \item[Fig.\  2  :]  The $r-p$ phase diagram for the
    BAW model with one and two offsprings with relative ratio
    (static branching). The full line between the Monte Carlo data
    is a guide to the eye.

   \item[Fig.\  3  :]  Plots of the effective exponents against
    $1/t$. Five curves from top to bottom
    in each panel correspond to $p=0.5080$, 0.5098, 0.5105,
    0.5112, 0.5140.
    Thick lines are critical lines ($p=0.5105$).
   \item[Fig.\  4  :]  The $r-p$ phase diagram for the
    BAW model with one and two offsprings with relative ratio
    (dynamic branching). The full line between the Monte Carlo data
    is a guide to the eye.

\end{description}

\end{document}